# Sources of cosmic microwave radiation and dark matter identified: millimeter black holes (m.b.h.)

Antonio Alfonso-Faus[1], Màrius Josep Fullana i Alfonso[2]

[1]*Department of Aerotecnia, E.U.I.T. Aeronáutica, Plaza Cardenal Cisneros s/n, 28040 Madrid, SPAIN*

[2]*Institut de Matemàtica Multidisciplinària, Universitat Politècnica de València, Camí de Vera s/n, 46022 València, SPAIN*

April 9, 2010

**Abstract.** - The universe is filled with blackbody millimeter radiation (CMBR), temperature 2.7º Kelvin[1]. Big-bang cosmology explains this by the initial thermalization of photons scattered by electrons[2]. This explanation requires *ad hoc* previous existence of photons and thermal electrons. On the other hand most of the mass of the universe is unknown *dark matter*[3]. It explains anomalous dynamical properties, like that of stars in galaxies[4, 5, 6]. Alternatively the anomalies have been explained by adjusting and modifying well known laws ("Modified Newtonian dynamics"[7]). Here we show that millimeter black holes (m.b.h.) explain both: the background radiation, by its partial "evaporation", and the dark matter. Black holes emit blackbody radiation (Hawking[8] evaporation), and this is what is observed in the CMBR. Millimeter size black holes emit blackbody radiation at a temperature of 2.7º Kelvin, and this is the resulting CMBR. Partial evaporation of ~10^30 m.b.h. gives the observed background field of photons being emitted and absorbed at the same rate by the m.b.h. The **number of photons is constant, as observed. Their temperature decreases with time because the mass of the m.b.h. (and therefore its size) increases with time (the mass-boom effect[9]). The total mass of the m.b.h. is the dark matter. Hence dark matter is not so "dark" after all. Two important cosmological items are here identified by only one source: millimeter black holes.**

1.- INTRODUCTION

The cosmic microwave background radiation (CMBR) was discovered by Penzias and Wilson[1] in 1965. It is blackbody radiation at a temperature of 2.7º K and fills all space being almost homogeneous and isotropic, to a very high degree. Its presence has been currently taken in support of the Big-Bang cosmological model. The CMBR, and the Hubble observation of the red shift interpreted as an expansion of the universe, are the two main pillars of this model. However, for the CMBR case, one needs to explain two things: the presence of photons (at least identifying their sources) and their thermalization by scattering electrons[2]. In the case of the red shift (Big-Bang expansion of the universe) this model is still in deep need of study (need to add different kinds of inflation, variable speed of light c, etc.). Several alternative models for the origin of CMBR and its blackbody spectrum have been developed from different point of views[10,11,12,13].

On the other hand, the existence of an unknown kind of gravitational dark matter has been proposed many years ago[3]. The dynamical gravitational effects observed in stars and galaxies[4,5,6] have not been explained by a direct observation of normal baryonic matter. The name "dark" is due to the fact that no one has seen it. The proposed existence of dark matter in the universe can be taken as an *ad hoc* postulate to fit the observed dynamics, and no one knows its constitution or properties. Certainly is not baryonic matter. The only evidence for dark matter is the gravitational action on stars and galaxies. The situation is similar to the one that occurred when observing the motion of planets in the solar system, many years ago. The existence of the furthest planets was predicted by their gravitational effect on the well known orbits of the closer ones. The difference here is that this prediction was confirmed by observation. In the case of the dark matter the prediction has not been confirmed so far by any other evidence than the gravitational action.

The dynamical anomalies observed in stars and galaxies could also be explained without invoking the presence of unknown dark matter. This is the case of the MOND hypothesis[7]: a theory of modified Newtonian dynamics by means of an *ad hoc* arrangement of the gravitational laws to fit the observed anomalous motions.

The picture of these two constituents of the universe, the CMBR and the dark matter, has one thing in common: the need of so far *ad hoc* assumptions, more than one, to explain the observations. We will substitute all of them by only one: the presence of millimeter black holes (m.b.h.). These are able to explain the source of the CMBR, its properties, and the dark matter issue.

2. - MILLIMETER BLACK HOLES

The CMBR is blackbody, the same kind of radiation as the photons emitted by a black hole (Hawking evaporation[8]). These black holes (b.h.) have a mass $m$ proportional to its size $r$ given by the Schwarzschild formula:

$$2\frac{Gm}{c^2} = r \qquad (1)$$

The typical wavelength λ of the photons in a blackbody distribution is determined by the blackbody temperature T. For the CMBR T ≈ 2.7° K, and for this value λ (Wien´s law) is very close to 0.1 cm. This explains the name of microwave or millimeter radiation. The point here is that the Hawking radiation from a b.h. has a blackbody distribution with a typical wavelength given by Wien's law. The mass *m* of a b.h. emitting with the same kind of radiation as it is observed in the CMBR, with T = 2.725° K is given by

$$m = \frac{\hbar c^3}{8\pi GkT} = 4.28 \times 10^{25} \; grams \qquad (2)$$

where $\hbar$ is the reduced Planck constant, *c* is the speed of light, *G* is the gravitational constant and *k* is the Boltzmann constant. The gravitational radius of this mass is about 0.00635 cm. The time that this black hole takes to dissipate is much larger than the present age of the Universe. Their size, proportional to their mass, increases due to the mass-boom effect[9]. Certainly only a partial evaporation has occurred as of today.

An estimate of the total mass of the universe can be obtained from the physical properties (length, mass and time) of the Planck´s quantum black hole. These physical properties are given by the simplest combination of G, $\hbar$ and c. They are also the result of equating the gravitational radius of its mass, $2Gm_p/c^2$, to its Compton wavelength, $\hbar/m_p c$ :

$$l_p = (\hbar 2G/c^3)^{1/2} = 2.285 \times 10^{-33} \; cm$$

$$m_p = (\hbar c/2G)^{1/2} = 1.539 \times 10^{-5} \; g \qquad (3)$$

$$t_p = (\hbar 2G/c^5)^{1/2} = 7.624 \times 10^{-44} \text{ sec}$$

Taking the present age of the universe to be $t = 1.37 \times 10^{10}$ years $= 4.32 \times 10^{17}$ sec, as current estimates. This gives a visible size of $R = ct \approx 1.3 \times 10^{28}$ cm. The case for the universe to be a quantum black hole has been analyzed elsewhere[14]. The scale factor F between the universe and the Planck´s quantum black hole is then from (3)

$$F \approx 1.3 \times 10^{28}/2.285 \times 10^{-33} \approx 5.69 \times 10^{60} \qquad (4)$$

Using this factor for the time and mass in (3) we get for the universe

$$F \cdot t_p = 5.69 \times 10^{60} \times 7.624 \times 10^{-44} \text{ sec} \approx 1.37 \times 10^{10} \text{ years}.$$

$$F \cdot m_p = 5.69 \times 10^{60} \times 1.539 \times 10^{-5} \text{ grams} \approx 8.76 \times 10^{55} \text{ grams} \qquad (5)$$

It is clear that the scale F fits the length, time and mass ratios between the two quantum black holes, as shown elsewhere[14]. Then, from (5) if $M \approx 8.76 \times 10^{55}$ grams is all the gravitational matter of the universe then we see that its size $R \approx 1.3 \times 10^{28}$ cm is also given by[15]

$$R \approx 2\frac{GM}{c^2} \approx 1.3 \times 10^{28} \, cm \qquad (6)$$

This important relation implies that the gravitational radius of the universe is in fact of the order of the product $ct \approx 1.3 \times 10^{28}$ cm. The number N of the partially evaporated m.b.h. of mass m given by (2), needed to explain the matter M of the universe is then given by

$$N = \frac{M}{m} \approx 2 \times 10^{30} \qquad (7)$$

The radius of the sphere that contains just one of these m.b.h. (assuming a uniform distribution all over the entire universe) is

$$\frac{R}{N^{1/3}} \approx 10^{18} \, cm \qquad (8)$$

The time for the photons emitted from each m.b.h. to fill this sphere (and therefore to have all the universe filled with blackbody radiation) is then about $3 \times 10^7$ sec. i.e. about one year. This means that the CMBR filled the universe when it had about one year of age, very much earlier than the current big-bang estimates.

We note that the m.b.h. establish and fill an intermediate scale between the two extremes: the Plank´s scale and the scale of the whole visible Universe. We can check that the product of the mass of the Universe times the Planck´s mass is equal to the square of the mass of the m.b.h.:

$$M_u \cdot m_p \approx (\text{mass of m.b.h.})^2 \qquad (9)$$

This duality between the Universe and the Planck´s scale can be equivalently presented as

$$\frac{M_u}{(\text{mass of m.b.h.})} \approx \frac{(\text{mass of m.b.h.})}{m_p} \qquad (10)$$

The scale between $M_u$ and $m_p$ is about $5.7 \times 10^{60}$. As we will see the scale between the Universe and the m.b.h. is the same as the scale between the m.b.h. and the Planck´s mass, i.e. about $(5.7 \times 10^{60})^{1/2} = 2 \times 10^{30}$. This is a very interesting numerology because the numerical numbers involved are very large. Therefore these expressions are expected not to be just a non-significant coincidence. The interesting point is that we get the same very large numbers from two apparently unrelated approaches.

3. – CONCLUSIONS

The origin of CMBR and dark matter can be explained with only one hypothesis. A homogeneous distribution of m.b.h. of 0.00635 cm could be the source of CMBR as the emission of such black holes. Also the blackbody spectrum of CMBR is deduced. The same m.b.h. will give the mass quantity which is stated as dark matter. Therefore, the effects explicated with dark matter could be stated with such black holes. The unknown nature of dark matter will then disappear. The distribution of such black holes would be the same postulated for dark matter. Finally, we identify the m.b.h. scale as the intermediate one, in a geometric mean sense, between Planck´s scale and the scale of the visible Universe.

**Correspondence** and requests for materials should be addressed to A. Alfonso-Faus, (antonio.alfonso@upm.es) or M.J. Fullana i Alfonso (mfullana@mat.upv.es).